\documentstyle[12pt]{article}

\def\thefootnote{\fnsymbol{footnote}}
\catcode`\@=11
\def\numberbysection{\@addtoreset{equation}{section}
        \def\theequation{\thesection.\arabic{equation}}}
\def\beq{\begin{equation}}
\def\eeq{\end{equation}}
\def\barr{\begin{eqnarray}}
\def\beqa{\begin{eqnarray}}
\def\earr{\end{eqnarray}}
\def\eeqa{\end{eqnarray}}
\def\winf{W_{1+\infty}\ }
\def\u1{\widehat{U(1)}}
\def\v{V\,}
\def\w{W\,}
\def\vb{{\overline V}\,}
\def\wb{{\overline W}\,}

\newcommand{\secn}[1]{Section~\ref{#1}}

\newcommand{\eq}[1]{Eq.~(\ref{#1})}

\newcommand{\nl}{\nonumber \\}

\renewcommand{\theequation}{\thesection.\arabic{equation}}
\newcommand{\EQ}{\begin{equation}}
\newcommand{\EN}{\end{equation}}
\newcommand{\bea}{\begin{eqnarray}}
\newcommand{\ena}{\end{eqnarray}}

\newcommand{\NP}[1]{{\it Nucl.\ Phys.\ }{\bf #1}}
\newcommand{\PL}[1]{{\it Phys.\ Lett.\ }{\bf #1}}

\newcommand{\CMP}[1]{{\it Comm.\ Math.\ Phys.\ }{\bf #1}}
\newcommand{\PR}[1]{{\it Phys.\ Rev.\ }{\bf #1}}
\newcommand{\PRL}[1]{{\it Phys.\ Rev.\ Lett.\ }{\bf #1}}

\newcommand{\IJMP}[1]{{\it Int.\ Jour.\ Mod.\ Phys.\ }{\bf #1}}
\renewcommand{\thefootnote}{\fnsymbol{footnote}}
\numberbysection

\begin{document}
\begin{titlepage}
\begin{center}
\hfill DFTT 13/96 \\
\hfill April 1996
\vskip 0.8cm
{\Large \bf The extended conformal theory of the 
Calogero-Sutherland model \footnote{Talk given at the 
{\it IV} Chia Meeting ``Common Trends in Condensed Matter
and High Energy Physics'', Chia Laguna, Cagliari, Italy,
~~3--10 Sep. 95.}}
\vskip 0.8cm
Raffaele~CARACCIOLO,~~Marialuisa~FRAU,~~Stefano~SCIUTO \\
\vskip 0.3cm
{\em      Dipartimento di Fisica Teorica,
          Universit\`a di Torino,\\
	     and I.N.F.N. Sezione di Torino \\
          Via P. Giuria 1, I-10125 Torino, Italy}
\vskip 0.6cm
Alberto LERDA \\
\vskip 0.3cm
{\em       Dipartimento di Scienze e Tecnologie Avanzate \footnote{
           II Facolt\'a di Scienze M.F.N., Universit\`a di Torino
           (sede di Alessandria), Italy.} and\\
           Dipartimento di Fisica Teorica, Universit\`a di Torino,\\
           and I.N.F.N., Sezione di Torino \\
           Via P.Giuria 1, I-10125 Torino, Italy}
\vskip 0.6cm
Guillermo~R.~ZEMBA \\
\vskip 0.3cm
{\em       Centro At\'omico Bariloche \\
           8400 - San Carlos de Bariloche ~(R\'{\i}o Negro),
Argentina}

\end{center}
\vskip 1cm
\begin{abstract}
\noindent
We describe the recently introduced method of Algebraic 
Bosonization of $(1+1)$-dimensional fermionic systems by
discussing the specific case of the Calogero-Sutherland model.
A comparison with the Bethe Ansatz results is also presented.
\end{abstract}
\vfill
\end{titlepage}
\pagenumbering{arabic}
\renewcommand{\thefootnote}{\arabic{footnote}}
\setcounter{footnote}{0}
\setcounter{page}{1}

\section{The effective theory}
\label{linear}
\bigskip

The Calogero-Sutherland model describes a system
of $N$ non-relativistic spinless fermions of mass $m$
moving on a circle
of length $L$ with a pairwise interaction proportional to the
inverse square of the chord distance between the
two particles~~\cite{cal,sut}. Denoting by $x_i$ the coordinate 
of the $i$-th fermion along the circle and choosing units such
that $\hbar=2m=1$, the hamiltonian reads
$$
h=-\sum_{j=1}^N \frac{\partial^2}{\partial x_j^2}
+g\,\frac{\pi^2}{L^2}\sum_{j<k}
\frac{1}{\sin^2 (\pi(x_j-x_k)/L) }~~~,
$$
where $g$ is the coupling constant. In the following we shall take $N$
odd, without any loss of generality.

This model is exactly solved by Bethe Ansatz and all its fundamental
properties can be obtained from this solution \cite{sut}. In particular,
the low-energy excitations above the ground state are
gapless, and thus the long-distance properties of the system are
described by a conformal field theory \cite{bpz,hald1,kaya}. 

The second quantized hamiltonian corresponding to $h$ is
the sum of the kinetic term
\beq
H_0 = \left(\frac{2\pi}{L}\right)^2\sum_{n=-\infty}^\infty
n^2\,\psi_n^{\dagger}\,\psi_n~~~,
\label{h0cs}
\eeq
and the interaction term \cite{clz}
\beq
H_I =- g \,\frac{\pi^2}{L^2} \sum_{l,n,m=-\infty}^\infty |l|
~\psi_{m+l}^{\dagger}\,\psi_{n-l}^{\dagger}\,
\psi_n \,\psi_m ~~~,
\label{hics}
\eeq
where $\psi_n$ are fermionic oscillators of momentum
$k_n=\left(2\pi/L\right)n$, satisfying standard anticommutation
relations.

The hamiltonian $H_0$ describes a free fermionic system whose
ground state is 
$$
|\Omega\rangle = \psi^\dagger_{-n_F} \dots \psi^\dagger_{n_F}
|0\rangle~~~~,
$$
with $n_F = (N-1)/2$ .
Only the oscillators near the Fermi points $\pm n_F$
play an important role in physical processes,
producing the low-energy excitations above $|\Omega\rangle$ 
and determining the large-distance properties
of the system encoded in the effective theory.
In order to write down the corresponding hamiltonian,
we define shifted fermionic operators associated 
to the small fluctuations around each Fermi point according to
\beq
a_r\equiv\psi_{n_F+r}~~~,~~~
b_r\equiv\psi_{-n_F-r}~~~~.
\label{arbr}
\eeq
The quantity $2\pi\left(r-1/2\right)/N$ ($-2\pi\left(r-1/2\right)/N$)
represents the momentum of the oscillator $a_r$ ($b_r$) relative to the
right (left) Fermi point.
The integer index $r$ is allowed to vary only in a
finite range between $-\Lambda_0$ and $+\Lambda_0$,
where $\Lambda_0$ is a bandwidth cut-off chosen such that
$\Lambda_0 \ll n_F$, and $\Lambda_0=o(n_F)=o(N)$ in the thermodynamic
limit $N\to\infty$.
Roughly speaking,
$\Lambda_0$ indicates how far from the Fermi points one can go
without leaving the effective regime. The two sets of oscillators 
$a$ and $b$ in \eq{arbr} define two independent and finite
branches of excitations around each Fermi point, such that
\bea
a_r|\Omega\rangle = 0~~~&,&~~~
b_r|\Omega\rangle = 0 ~~~~{\rm for}~~r=1,2,\ldots,\Lambda_0~~~,
\nl
a_s^\dagger|\Omega\rangle = 0~~~&,&~~~
b_s^\dagger|\Omega\rangle = 0 ~~~~{\rm for}~~
s=0,-1,-2,\ldots,-\Lambda_0~~~.
\label{abvac1}
\ena
The kinetic part of the hamiltonian for these effective degrees of
freedom reads
\beq
{\cal H}_0=
\left(\frac{2\pi}{L}\right)^2\
\sum_{r=-\Lambda_0}^{\Lambda_0}\
\left(n_F+r\right)^2 \,
\left(:a^\dagger_r\,a_r:+:b^\dagger_r\,b_r:\right)~~~.
\label{kinab}
\eeq

Let us now consider the interaction hamiltonian $H_I$.
We seek the corresponding effective operator
${\cal H}_I$ in terms of bilinear fermionic forms, which
can be naturally interpreted within
the algebraic context of extended conformal theories.
To this end, a reordering of the oscillators in \eq{hics}
is needed. However, this causes the appearance of a divergent 
two-fermion term, and thus it is necessary to introduce a 
regularization prescription to give a meaning to our formulas.
We perform a ``periodic regularization'', by dividing the momentum 
space into fictitious Brillouin zones of amplitude $2M$, and 
identifying $l$ and $l'$ if $l'=l+2M\,k$ for any integer $k$.
The arbitrary number $M$ has the only constraint $M \gg n_F$,
such that the physical region of interest is inside the first zone.
Thus, we can calculate the effect of the small oscillations 
around the Fermi points $\pm n_F$ in the first Brillouin zone,
and, at the end, let $M\to\infty$ to recover the original continuum
theory. It can be shown that this procedure
is consistent and leads to finite and meaningful results \cite{flsz}.
With this regularization the effective hamiltonian 
${\cal H}_I(M)$ is
$$
{\cal H}_I(M)={\cal H}_{forw}(M)+ {\cal H}_{back}(M)
+ {\cal H}'(M)
$$
where the forward scattering part ${\cal H}_{forw}(M)$ is
\bea
{\cal H}_{forw}(M) &=& \!- g\,\frac{\pi^2}{L^2}~
{ \sum_\ell}' \sum_{r,s=-\Lambda_0}^{\Lambda_0} \big|\ell\big|_M
\left[:a^\dagger_{r-\ell}\,a_r:\,:a^\dagger_{s+\ell}\,a_s:\right.
\label{hifcs}\\
&& \!+ \left. :b^\dagger_{r+\ell}\,b_r:\,
:b^\dagger_{s-\ell}\,b_s:
+:a^\dagger_{r-\ell}\,a_r:\,:b^\dagger_{s-\ell}\,b_s:
+:b^\dagger_{r+\ell}\,b_r:\,:a^\dagger_{s+\ell}\,a_s:\right]
\nonumber
\ena
with $\big|\ell\big|_M$ being the periodic extension modulo $2M$
of the absolute value, defined by
$\big|\kappa\big|_M ~=~ |\kappa|~~,~~
\big|M+\kappa\big|_M ~=~ \big|-M+\kappa\big|_M=M-|\kappa|$
for $|\kappa|\leq M$.
The symbol $\,\sum'$ means that the sum over $\ell$
is restricted to values satisfying the constraints:
\bea
|r|\leq \Lambda_0 ~~~&,&~~~ |s|\leq \Lambda_0 ~~~,\nl
|r\mp\ell|\leq \Lambda_0~~~&,&~~~|s\pm\ell|\leq \Lambda_0~~~.
\nonumber
\ena
\noindent
The backward scattering hamiltonian, ${\cal H}_{back}(M)$, is
\bea
{\cal H}_{back}(M)&=& -g\,\frac{\pi^2}{L^2}~
{ \sum_\ell}' \sum_{r,s=-\Lambda_0}^{\Lambda_0}
 \left[
\big|2n_F+\ell \big|_M\,
b^\dagger_{-r+\ell}\,a_r\,a^\dagger_{-s+\ell}\,b_s \right. \nl
&& + \left.\big|-2n_F+\ell \big|_M\,
a^\dagger_{-r-\ell}\,b_r\,b^\dagger_{-s-\ell}\,a_s \right]~~~,
\label{hibcs}
\ena
while the term ${\cal H}'$, arising from a reordering of the
oscillators in \eq{hics},
is
\beq
{\cal H}'(M) = g\,\frac{\pi^2}{L^2}\,M^2\sum_{r=-\Lambda_0}^{\Lambda_0}
\left(:a^{\dagger}_r\, a_r :+:b^{\dagger}_r\, b_r:\right)
~~~.
\label{hidue}
\eeq

The effective theory is defined by letting $\Lambda_0\to\infty$, 
$N\to\infty$ and keeping $\Lambda_0 \ll N$. 
This avoids the introduction of spurious low-energy states 
(the $O(1/N^2)$-corrections bend the dispersion curve) \cite{flsz}.
 For ease of notation, when $\Lambda_0$ and $N \to \infty$,
we write the free effective hamiltonian
${\cal H}_0$ simply as in \eq{kinab}
with all sums extended from $-\infty$ to $+\infty$, and
the limit $N\to\infty$ left implicit.
Even though ${\cal H}_0$ now contains
new oscillators, it still
acts only on low-energy states
with particle and hole momenta bounded by $\Lambda_0\ll N$.
Nonetheless, the extension of the dispersion curve to
infinity is not free of consequences. In fact,
since now  Eqs. (\ref{abvac1}) hold for {\it any} integers
$r$ and $s$, the ground state $|\Omega\rangle$ corresponds
to the surface of two {\it infinite}
 Fermi seas.

Using the definition of $n_F$, we easily
see that \eq{kinab} becomes
\beq
{\cal H}_0=\left(2\pi\rho_0\right)^2 \sum_{r=-\infty}^\infty
\left[
\frac{1}{4} + \frac{1}{N} \left(r-\frac{1}{2}\right) +
\frac{1}{N^2}\left(r-\frac{1}{2}\right)^2 \right]
\left(:a^{\dagger}_r \,a_r:+:b^{\dagger}_r \,b_r:\right)\ ,
\label{kindue}
\eeq
where $\rho_0 =N/L$ is the density, which is kept fixed in the
thermodynamic limit.
Analogously, the forward scattering terms, \eq{hifcs} (which 
are actually independent of the regularization parameter $M$),
simply turn into
\bea
{\cal H}_{forw} &=& -\frac{g}{4}\left(2\pi \rho_0\right)^2
\frac{1}{N^2}
\sum_{\ell,r,s=-\infty}^\infty|\ell|
\left[:a^\dagger_{r-\ell}\,a_r:\,:a^\dagger_{s+\ell}\,a_s:\right.
\nl
&& + \left. :b^\dagger_{r+\ell}\,b_r:\,:b^\dagger_{s-\ell}\,b_s:
+2\,:a^\dagger_{r-\ell}\,a_r:\,:b^\dagger_{s-\ell}\,b_s: \right]~~~.
\label{hif1cs}
\ena
\noindent
The backward scattering terms, instead, explicitly depend on $M$.
Rewriting \eq{hibcs} in terms of fermion
bilinears and removing the bandwidth cut-off leads to
the appearance of a divergent two-body term,
in addition to a finite four-fermion piece.
However, after combining these terms with ${\cal H}'(M)$,
the resulting hamiltonian in \eq{hidue} is finite. 
In fact \cite{flsz}
\bea
&&\lim_{M \to \infty} \left({\cal H}_{back}(M) +
{\cal H}'(M)\right)
\nl
&=& \frac{g}{2}\left(2\pi \rho_0\right)^2
\frac{1}{N^2}\sum_{\ell,r,s=-\infty}^\infty
(N+r+s-\ell-1)
\,:a^\dagger_{r-\ell}\,a_r:\,
:b^\dagger_{s-\ell}\,b_s:
\label{hdos}\\
&& + \frac{g}{4}\left(2\pi \rho_0\right)^2
\frac{1}{N^2}
\sum_{r=-\infty}^\infty \left[N^2 + N(2r-1) + r(r-1) \right]
\left( : a^{\dagger}_r \,a_r:+
:b^{\dagger}_r\, b_r:\right)~~~.
\nonumber
\ena

We now summarize our results by writing the complete
effective hamiltonian of the model, which is
\beq
{\cal H} = \left(2\pi \rho_0\right)^2
\sum_{k=0}^2 \frac{1}{N^k}\,{\cal H}_{(k)}
\label{seriesc}
\eeq
where
\beq
{\cal H}_{(0)} =\frac{1}{4}\,(1+g)\sum_{r=-\infty}^\infty  
\left( : a^{\dagger}_r \,a_r:+
:b^{\dagger}_r\, b_r:\right)~~~,
\label{hc0}
\eeq
\bea
{\cal H}_{(1)} &=& \left(1+\frac{g}{2}\right)
\sum_{r=-\infty}^\infty  \left(r-\frac{1}{2}\right)
\left( : a^{\dagger}_r \,a_r:+
:b^{\dagger}_r\, b_r:\right)\nl
&& + \frac{g}{2}\sum_{\ell,r,s=-\infty}^\infty
:a^\dagger_{r-\ell}\,a_r:\,
:b^\dagger_{s-\ell}\,b_s:~~~,
\label{hc1}
\ena
and
\bea
{\cal H}_{(2)} &=& \sum_{r=-\infty}^\infty  
\left[\left(r-\frac{1}{2}\right)^2 + 
\frac{g}{4}\left(r^2-r\right)\right]
\left( : a^{\dagger}_r \,a_r:+
:b^{\dagger}_r\, b_r:\right)\nl
&& - \frac{g}{4} \sum_{\ell,r,s=-\infty}^\infty|\ell|
\left[:a^\dagger_{r-\ell}\,a_r:\,:a^\dagger_{s+\ell}\,a_s:
+:b^\dagger_{r+\ell}\,b_r:\,:b^\dagger_{s-\ell}\,b_s: \right.\nl
&&+\left.2\,:a^\dagger_{r-\ell}\,a_r:\,:b^\dagger_{s-\ell}\,b_s: \right]
\label{hc2}\\
&& + \frac{g}{2} \sum_{\ell,r,s=-\infty}^\infty(r+s-\ell-1)
:a^\dagger_{r-\ell}\,a_r:\,:b^\dagger_{s-\ell}\,b_s: ~~~.
\nonumber
\ena
Notice that there are no contributions to ${\cal H}$ to order $1/N^3$
and higher.
Note also that, despite the overall factor $1/L^2$ in \eq{hibcs}, 
the backward scattering part contributes to both the zeroth- 
and first-order effective hamiltonians ${\cal H}_{(0)}$ and 
${\cal H}_{(1)}$, due to normal ordering effects. On the contrary,
the forward scattering part, \eq{hif1cs}, 
contributes only to ${\cal H}_{(2)}$.

\vskip 1.5cm
\section{\bf The $\winf$structure of the model}
\label{csw}

We now show that the effective hamiltonian ${\cal H}$ in 
\eq{seriesc} can be given an elegant algebraic 
interpretation.
In fact, if we introduce the fermionic bilinear operators
\barr
\v^0_\ell &=& \sum_{r=-\infty}^{\infty}: a^\dagger_{r-\ell}\, a_r :
~~~,\nl
\v^1_\ell &=& \sum_{r=-\infty}^{\infty} \left(\,r-{\ell+1\over 2}\,
\right) : a^\dagger_{r-\ell}\, a_r :~~~,\label{fockw}\\
\v^2_\ell &=& \sum_{r=-\infty}^{\infty} \left(\,r^2 -(\ell+1)\ r +
{{(\ell+1)(\ell+2)}\over 6}\,
\right) :  a^\dagger_{r-\ell}\, a_r :~~~,
\nonumber
\earr
and $\vb_\ell^0$, $\vb_\ell^1$ and $\vb_\ell^2$ defined as above
with $: a^\dagger_{r-\ell}\, a_r :$ replaced by
$: b^\dagger_{r-\ell}\, b_r :$, then the operators
${\cal H}_{(k)}$ of Eqs. (\ref{hc0})-(\ref{hc2}) become respectively
\beq
{\cal H}_{(0)} =  \frac{1}{4} (1+g) \left(\v^0_0+
\vb^0_0\right)~~~,
\label{hcv0}
\eeq
\beq
{\cal H}_{(1)} = \left(1+\frac{g}{2}\right) \left(\v^1_0+
\vb^1_0\right)+\frac{g}{2} \sum_{\ell=-\infty}^{\infty}
\v^0_{\ell}\, \vb^0_{\ell}~~~,
\label{hcv1}
\eeq
\bea
{\cal H}_{(2)} &=&
\left(1+\frac{g}{4}\right) \left(\v^2_0+\vb^2_0\right)
-\frac{1}{12} (1+g) \left(\v^0_0+\vb^0_0\right) \nl
&& - \frac{g}{4}\sum_{\ell=-\infty}^{\infty} |\ell |
\left( \v^0_{\ell}\,\v^0_{-\ell}+
\vb^0_{-\ell} \,\vb^0_{\ell}+
2\,\v^0_{\ell} \,\vb^0_{\ell}\right)\nl
&& + \frac{g}{2}
\sum_{\ell=-\infty}^{\infty} \left( \v^1_{\ell}\,\vb^0_{\ell}+
\v^0_{\ell} \,\vb^1_{\ell} \right) ~~~.
\label{hcv2}
\ena

The reason for this rewriting lies in the fact that the operators
in Eqs. (\ref{fockw}) are the lowest generators of the 
infinite dimensional algebra known
as $\winf$ \cite{shen,kac1}. Its general form is
\beq
\left[\ V^i_\ell, V^j_m\ \right] = (j\ell-im) V^{i+j-1}_{\ell+m}
+q(i,j,\ell,m)V^{i+j-3}_{\ell+m}
+\cdots +\delta^{ij}\delta_{\ell+m,0}\ c\ d(i,\ell) \ ,
\label{walg}
\eeq
where the structure constants $q(i,j,\ell,m)$ and $d(i,\ell)$ 
are polynomial in their arguments, $c$ is the central charge, 
and the dots denote a finite number of terms involving the operators 
$V^{i+j-2k}_{\ell+m}\ $.
In our case $c=1$, and the relevant commutation relations are
\barr
\left[\ V^0_\ell,V^0_m\ \right] & = &  \ell\ \delta_{\ell+m,0} ~~~,
\label{walg0} \\
\left[\ V^1_\ell, V^0_m\ \right] & = & -m\ V^0_{\ell+m} ~~~,
\label{walg01}\\
\left[\ V^1_\ell, V^1_m\ \right] & = & (\ell-m)V^1_{\ell+m} + 
\frac{1}{12}\ell(\ell^2-1) \delta_{\ell+m,0}~~~,
\label{walg1}\\
\left[\ V^2_\ell, V^0_m\ \right] &=& -2m\ V^1_{\ell+m}~~~,
\label{walg20}\\
\left[\ V^2_\ell, V^1_m\ \right] &=& (\ell-2m)\ V^2_{\ell+m} -
   \frac{1}{6}\left(m^3-m\right) V^0_{\ell+m}~~~.
\label{walg21}
\earr
Eqs. (\ref{walg0}) and (\ref{walg1}) show respectively
that the generators $V^0_\ell$ satisfy the 
Abelian Kac-Moody algebra $\u1$, and the generators $V^1_\ell$ 
satisfy the Virasoro algebra.
The operators $\vb^i_\ell$ close the same algebra (\ref{walg}) 
and commute with the $\v^i_\ell$'s. 

The $c=1$ $\winf$algebra can be also realized by {\it bosonic} 
operators, through a generalized Sugawara construction \cite{kac1}.
In fact, if one introduces the right and left moving modes,
$\alpha_\ell$ and ${\overline \alpha}_\ell$, of a free compactified 
boson, one can check that the commutation relations (\ref{walg})
are satisfied by defining $\v^i_\ell$ (we only write the expressions
for $i=0,1,2$) as
\barr
\v^0_\ell &=& \alpha_\ell ~~~,
\label{mod0}\\
\v^1_\ell &=& {\frac{1}{2}} \sum_{r= -\infty}^{\infty}
:\, \alpha_{r}\,\alpha_{\ell-r}\,
:~~~,\label{mod1}\\
\v^2_\ell &=& {\frac{1}{3}} \sum_{r, s = -\infty}^{\infty}
:\, \alpha_{r}\,\alpha_s\, \alpha_{\ell-r-s}\,:~~~,
\label{mod2}
\earr
and analogously the $\vb^i_\ell$ in terms of 
${\overline \alpha}_\ell$.

The major advantage for choosing the basis of the $\winf \times 
{\overline \winf}$ operators is that, once the algebraic content 
of the theory has been established in the free fermionic picture,
other bosonic realizations of the {\it same} algebra can be used,
and the free value of the compactification radius of the boson
can be chosen to diagonalize the total hamiltonian.
This is the reason for calling this procedure
{\it algebraic bosonization} \cite{flsz}.

In the fermionic description it is easy to see that the highest 
weight states of the $\winf \times {\overline \winf}$ algebra
are obtained by adding $\Delta N$ particles to the 
ground state $|\Omega \rangle$, and by moving $\Delta D$ 
particles from the left to the right Fermi point; they are 
denoted by $|\Delta N,\Delta D\rangle_0$.
The descendant states, 
$$
|\Delta N , \Delta D ; \{k_i\},\{{\overline k}_j\} 
\rangle_0 \ = 
\v^0_{-k_1} \dots \v^0_{-k_r} \vb^0_{-{\overline k}_1}
\dots \vb^0_{-{\overline k}_s}
|\Delta N , \Delta D \rangle_0~~~~,
$$
with $k_1 \ge k_2 \ge \dots \ge k_r > 0$, and
${\overline k}_1 \ge {\overline k}_2 \ge \dots 
\ge {\overline k}_s > 0$,
coincide with the particle-hole excitations obtained from
$|\Delta N , \Delta D \rangle_0$.
Using the expressions of $\v_0^0$ and $\vb_0^0$ given in
\eq{fockw}, one finds that the charges associated to 
these states are
\bea
\v_0^0 ~|\Delta N , \Delta D ; \{k_i\},\{{\overline k}_j\} 
\rangle_0&=& \left(\frac{\Delta N}{2} \ + \Delta D  \right)
|\Delta N , \Delta D ; \{k_i\},\{{\overline k}_j\} \rangle_0~~~,
\nl
\vb_0^0 ~ |\Delta N , \Delta D ; \{k_i\},\{{\overline k}_j\} 
\rangle_0 \ &=& \left(\frac{\Delta N}{2} \ - \Delta D  \right)
|\Delta N , \Delta D ; \{k_i\},\{{\overline k}_j\} \rangle_0~~~.
\nonumber
\ena
 From these eigenvalues, we clearly see that the bosonic field
built out of the $\v^0_\ell$ and $\vb_\ell^0$
(see \eq{mod0}) is compactified on a circle of radius $r_0=1$. 
This field describes the density fluctuations of the original 
free fermions.

Let us now consider the $1/N$-term of the effective
hamiltonian in \eq{hcv1}.
Due to the left-right mixing term proportional to $g$, 
${\cal H}_{(1)}$ is not diagonal on the states
$|\Delta N , \Delta D ; \{k_i\},\{{\overline k}_j\} \rangle_0$.
However, it can be diagonalized \cite{lutt} using the Sugawara 
construction: indeed, replacing $\left(\v_0^1 +\vb_0^1\right)$ 
with the expression given by \eq{mod1}, we obtain a quadratic 
form in $\v_\ell^0$ and $\vb^0_{\ell}$, 
which can be now diagonalized by means of the following 
Bogoliubov transformation
\barr
\w^0_{\ell}&=&\v^0_{\ell}\ \cosh \beta + \vb^0_{-\ell}\
\sinh \beta ~~~, \nl
\wb^0_{\ell}&=&\v^0_{-\ell}\ \sinh \beta +
\vb^0_{\ell}\ \cosh \beta
\label{bogo}
\earr
for all $\ell$, with $ \tanh 2\beta =g/(2+g)$.
Under this transformation ${\cal H}_{(1)}$ becomes
(up to an irrelevant additive constant)
\beq
{\cal H}_{(1)} = \frac{\lambda}{2}\left[\left(\w_0^0\right)^2
+\left(\wb_0^0\right)^2\right]
+\lambda\,\sum_{\ell=1}^\infty
\left(\w_{-\ell}^0\,\w_{\ell}^0+\wb_{-\ell}^0\,\wb_{\ell}^0\right)~~~,
\label{hw}
\eeq
where 
\beq
\lambda \equiv \exp(2\beta)= \sqrt{1+g}~~~~.
\label{deflam}
\eeq
In writing \eq{hw} we have used the property, implied by \eq{bogo},
that $\w^0_{\ell}$ and $\wb^0_{\ell}$
satisfy the same abelian Kac-Moody algebra with
central charge $c=1$ as the original
operators $\v^0_{\ell}$ and $\vb^0_{\ell}$ (cf. \eq{walg0}).
Then, by means of the generalized Sugawara construction,
we can define a new realization of the $\winf$algebra whose 
generators $\w_\ell^i$ and $\wb_\ell^i$ are forms of degree
$(i+1)$ in $\w^0_{\ell}$ and $\wb^0_{\ell}$ respectively.
Consequently, the effective hamiltonian,
up to order $1/N$, becomes
\cite{flsz}
\bea
{\cal H}_{(1/N)}&\equiv& \left(2\pi \rho_0\right)^2
\left( {\cal H}_{(0)}+\frac{1}{N}\,{\cal H}_{(1)}\right)
\nl
&=& \left(
2\pi\rho_0  \sqrt{\lambda}\right)^2
\left[\left(\frac{\sqrt{\lambda}}{4}\,\w_0^0
+\frac{1}{N}\,\w_0^1\right)+
\left(\,W~\leftrightarrow~{\overline W}\,\right)
\right]~~~,
\label{h1n}
\ena
and exhibits a left-right factorization in the {\it new } realization
of the $\winf$algebra.

Notice that ${\cal H}_{(1/N)}$ is not diagonal on
$|\Delta N , \Delta D ; \{k_i\},\{{\overline k}_j\} \rangle_0$,
because the highest weight states of the
new algebra do not coincide with the
vectors $|\Delta N , \Delta D \rangle_0$, as is clear from \eq{bogo}.
However, the Bogoliubov transformation does not mix states
belonging to different Verma moduli.
This implies that the new highest weight vectors,
$|\Delta N ; \Delta D \rangle_W$, are still
characterized by the numbers $\Delta N$ and $\Delta D$ with the
same meaning as before, but their charges are different.
More precisely
\bea
\w_0^0 ~|\Delta N ; \Delta D \rangle_W  &=&
\left(\sqrt{\lambda}\,\frac{\Delta N}{2}+
\frac{\Delta D}{\sqrt{\lambda}}\right)
|\Delta N ; \Delta D \rangle_W \nl
\wb_0^0 ~|\Delta N ; \Delta D \rangle_W  &=&
\left(\sqrt{\lambda}\,\frac{\Delta N}{2}-
\frac{\Delta D}{\sqrt{\lambda}}\right)
|\Delta N ; \Delta D \rangle_W~~~.
\label{vdo}
\ena
 From the last two equations we deduce that $\w^0_\ell$
and $\wb_\ell^0$ are still the modes of a compactified
bosonic field, but on a circle of radius 
$r=1/\sqrt{\lambda}=\exp(-\beta)$.
This new field describes the density fluctuations
of the {\it interacting} fermions.

The highest weight states $|\Delta N , \Delta D \rangle_W$
together with their descendants, denoted by
$|\Delta N , \Delta D ; \{k_i\},\{{\overline k}_j\} \rangle_W$,
form a {\it new} bosonic basis for our
theory that has no simple expression in
terms of the original free fermionic degrees of freedom.
The main property of this new basis is that it diagonalizes
the effective hamiltonian up to order $1/N$.
In fact, using Eqs. (\ref{h1n}) and (\ref{vdo}),
it is easy to check that
$$
{\cal H}_{(1/N)}
~|\Delta N , \Delta D ; \{k_i\},\{{\overline k}_j\} \rangle_W =
{\cal E}_{(1/N)}
~|\Delta N , \Delta D ; \{k_i\},\{{\overline k}_j\} \rangle_W
$$
where \cite{kaya}
\bea
{\cal E}_{(1/N)}&=& \left(
2\pi\rho_0  \sqrt{\lambda}\right)^2\Bigg[\frac{\lambda}{4}\,\Delta N +
\frac{1}{N}\left(\lambda\,\frac{\left(\Delta N\right)^2}{4}  +
\frac{\left(\Delta D\right)^2}{\lambda} +
k+{\overline k} \right)\Bigg]
\nl
&=& \mu \,\Delta N + \frac{2\pi v}{L}
\left(\lambda\,\frac{\left(\Delta N\right)^2}{4}  +
\frac{\left(\Delta D\right)^2}{\lambda} +
k+{\overline k} \right) ~~~,
\label{fsize}
\earr
with $k=\sum_i k_i$ and 
${\overline k}=\sum_j {\overline k}_j~$.
In the last line of \eq{fsize}, $\mu$ is the chemical potential 
and $v$ the Fermi velocity.
These eigenvalues are clearly degenerate when $k \geq 2$ or
${\overline k} \geq 2$.
Examining their structure one concludes that the 
Calogero-Sutherland interaction induces,
up to order $1/N$, three physical effects: a rescaling of the
chemical potential
\beq
\mu_0=\left(2\pi\rho_0  \right)^2/4~\longrightarrow~
\mu= \lambda^2\,\mu_0~~~~,
\label{chempot}
\eeq 
a rescaling of the Fermi velocity of the particles
\beq
v_0=2\pi\rho_0~\longrightarrow~ v= \lambda \,v_0~~~~,
\label{fervel}
\eeq
a change in the compactification radius of the bosonic field describing
the fermion density fluctuations
\beq
r_0=1~\longrightarrow~ r=r_0/{\sqrt{\lambda}}~~~~.
\label{comprad}
\eeq
Note that these effects have their origin in the backward 
scattering processes, which are the only interactions that 
contribute to the effective hamiltonian to order $1/N$.
In particular, to lowest order in $g$, the change in the 
compactification radius is induced by the left-right 
mixing terms of ${\cal H}_{(1)}$. 
Such terms are of the generic form $\v^0_.\,\vb^0_.\,$, 
and have conformal dimension $(1,1)$. 
Therefore, they are marginal operators, which cannot 
destroy the conformal symmetry of the free theory, but only 
change the realization of the conformal algebra \cite{bpz}.
In fact, these operators drive the theory out from the free 
realization in terms of $\v_\ell^0$ and $\vb_\ell^0$ to the 
interacting realization in terms of $\w_\ell^0$ and $\wb_\ell^0$. 
In this flow, the central charge of the conformal algebra 
remains unchanged, while the compactification
radius of the bosonic field scales as indicated.

The rescalings of the chemical potential and the
 Fermi velocity are, instead, a normal ordering effect:
to lowest order in $g$, they originate from
the left and right diagonal terms proportional to
$g$ in ${\cal H}_{(0)}$ and ${\cal H}_{(1)}$, which
arise from the two-body part of
the backscattering hamiltonian in (\ref{hdos}).

It is remarkable that despite their different origins,
these rescalings are characterized by only one function 
of the coupling constant: the parameter $\lambda$.
This fact implies the existence of relations among
$r$,$\mu$ and $v$; for example, one has 
$$v_0\,r_0^2 = v\,r^2~~~~,$$ which is typical 
of the Luttinger model \cite{lutt,solyo,hald}. 
It actually holds for all systems
with hamiltonian of the form (\ref{h1n}) to order $1/N$.

We should keep in mind that the derivation of the effective 
theory, as presented in \secn{linear}, is strictly 
perturbative; thus, an expansion in the coupling constant $g$
should be understood in all previous formulae.
However, if we limit our analysis to the $1/N$-terms,
nothing prevents us from improving our results by
extending them to all orders in $g$. Indeed, the Bogoliubov 
transformation (\ref{bogo}) diagonalizes the hamiltonian
${\cal H}_{(1)}$ {\it exactly}, and the resulting expression 
depends on the coupling constant only through $\lambda$, 
which contains all powers of $g$! This improvement is
a well-known result in the Luttinger model \cite{lutt,solyo},
but we would like to stress that in our case
it can be done only if we disregard
the $O(1/N^2)$-terms of the hamiltonian, because
the Bogoliubov transformation
(\ref{bogo}) does not diagonalize ${\cal H}_{(2)}$.

To investigate this issue, let us analyze the $1/N^2$-term 
of the effective hamiltonian \eq{hcv2}. Using the generalized 
Sugawara construction, Eqs. (\ref{mod0})-(\ref{mod2}), we 
first rewrite ${\cal H}_{(2)}$ as a cubic form in $\v^0_{\ell}$
and $\vb^0_{\ell}$, and then perform the Bogoliubov 
transformation (\ref{bogo}) to re-express it in terms of 
the $\w^i_\ell$ and $\wb^i_\ell$ generators. 
A straightforward calculation leads to
${\cal H}_{(2)} = {\cal H}_{(2)}' \ +\ {\cal H}_{(2)}''$
where
\bea
{\cal H}_{(2)}' &=&
\sqrt{\lambda} \left( \w^2_0+\wb^2_0 \right)-
\frac{\sqrt{\lambda^3}}{12}\left(\w_0^0+\wb_0^0\right)
\nl
&& - \frac{g}{2 \lambda}
\,\sum_{\ell=1}^{\infty}\,\ell\left(
\w^0_{-\ell}\,\w^0_{\ell}+ \wb^0_{-\ell}\,\wb^0_{\ell}
\right) ~~~,
\label{h2w'}
\ena
and
\beq
{\cal H}_{(2)}'' \,=\, -\,\frac{g}{2 \lambda}
\,\sum_{\ell=1}^{\infty}\,\ell\left(
\w^0_{\ell}\,\wb^0_{\ell}+ \w^0_{-\ell}\,\wb^0_{-\ell}
\right)~~~.
\label{h2w''}
\eeq
Neither ${\cal H}_{(2)}'$ nor ${\cal H}_{(2)}''$ are diagonal 
on the states 
$|\Delta N , \Delta D ; \{k_i\},\{{\overline k}_j\} \rangle_W$
considered so far. In fact, these states are
not in general eigenstates of $\left(\w_0^2+\wb_0^2\right)$,
and hence cannot be eigenstates of
${\cal H}_{(2)}'$; moreover, since they have definite
values of $k$ and ${\overline k}$, they cannot be eigenstates
of ${\cal H}_{(2)}''$ either, because this operator
mixes the left and right sectors.

It is not difficult, however, to overcome
these problems. Since
${\cal H}_{(2)}'$ and ${\cal H}_{(1/N)}$ commute with each
other, it is always possible to find suitable
combinations of the states
$|\Delta N , \Delta D ; \{k_i\},\{{\overline k}_j\} \rangle_W$
with fixed $k$ and ${\overline k}$ that diagonalize
simultaneously ${\cal H}_{(2)}'$ and ${\cal H}_{(1/N)}$,
therefore lifting the degeneracy of the
spectrum present to order $1/N$. 

The term ${\cal H}_{(2)}''$, instead, has to be treated 
perturbatively, but only to first order in $g$.
In fact to higher orders, the spurious states
introduced with the limit $\Lambda_0 \to \infty$
would also contribute as intermediate states.
These contributions, however, would be meaningless because 
the hamiltonian to order $O(1/N^2)$ is not even bounded below.
 From \eq{h2w''} it is easy to check that ${\cal H}_{(2)}''$ 
has vanishing expectation value on any state that is
simultaneously eigenstate of ${\cal H}_{(1/N)}$ and 
${\cal H}_{(2)}'$.
Thus, according to (non-degenerate) perturbation theory,
${\cal H}_{(2)}''$ has no effect on the energy spectrum
to first order in $g$.

In view of these considerations, we neglect
${\cal H}_{(2)}''$ and regard as the effective
hamiltonian the following operator
\bea
{\cal H}_{CS} &\equiv&
{\cal H}_{(1/N)} + \left(
2\pi\rho_0 \right)^2
\frac{1}{N^2}\,{\cal H}_{(2)}'\nl
&=&\left(
2\pi\rho_0  \sqrt{\lambda}\right)^2
\left\{\left[\frac{\sqrt{\lambda}}{4}\,\w_0^0
+\frac{1}{N}\,\w_0^1+
\frac{1}{N^2}\left(\frac{1}{\sqrt{\lambda}}\,\w_0^2
-\frac{\sqrt{\lambda}}{12}\,\w_0^0 \right.\right.
\right.\nl
&& -\left.\left.\left.
\frac{g}{2\lambda^2}\,\sum_{\ell=1}^\infty
\,\ell~\w_{-\ell}^0\,\w_\ell^0\right)
\right]+\left(\,W~\leftrightarrow~{\overline W}\,\right)
\right\}~~~.
\label{hcsf}
\ena
Obviously, to be consistent with our perturbative approach,
we should keep in the r.h.s. of \eq{hcsf} only the terms
that are linear in $g$.

We now compare the eigenvalues of 
${\cal H}_{CS}$ to the exact low-energy spectrum of the 
Calogero-Sutherland model obtained from the Bethe Ansatz 
method \cite{sut,kaya}.
Any low-energy solution of the Bethe Ansatz equations \cite{kore}
is labeled by a set of integer numbers 
\beq
I_j\ =\ \frac{2j-1-N'}{2} + \Delta D - {\overline n}_j +  n_{N'-j+1}
\label{Ii}
\eeq
where $N'= N + \Delta N$, and $j=1, \dots, N'$.
The integers $n_j$ are ordered according to 
$n_1\geq n_2\geq \dots \geq 0$
and are different from zero only if $j \ll N$ (and analogously 
for the ${\overline n}_j$).

By generalizing to order $1/N^2$ the procedure presented in 
Ref. \cite{kaya}, we have derived \cite{flsz} the exact energy of 
the excitation described by the numbers (\ref{Ii}):
\bea
\tilde{\cal E} &=& \left(2\pi\rho_0 \sqrt{\xi}\right)^2
\Bigg\{\Bigg[\frac{\sqrt{\xi}}{4}\,Q+\frac{1}{N}
\Bigg(\frac{1}{2}\,Q^2+ \sum_j n_j\Bigg)
\nl
&&+ \frac{1}{N^2}\Bigg(\frac{1}{3\sqrt{\xi}}\,Q^3
-\frac{\sqrt{\xi}}{12}\,Q +
\frac{2 \sum_j n_j}{\sqrt{\xi}}\,Q + \frac{\sum_j n_j^2}{\xi}
-\sum_j \left(2j-1\right) n_j\Bigg)\Bigg] 
\nl
&& +\left(Q\, \leftrightarrow \, {\overline Q}~,
{}~\{n_j\} \, \leftrightarrow \, 
\{{\overline n}_j\} \right)\Bigg\}~~~,
\label{eba}
\ena
where 
\beq
\xi=\frac{1+\sqrt{1+2g}}{2}~~~,
\label{xi}
\eeq
and
\beq
Q=\sqrt{\xi}\,\frac{\Delta N}{2}+
\frac{\Delta D}{\sqrt{\xi}}~~~~,~~~~
{\overline Q}=\sqrt{\xi}\,\frac{\Delta N}{2}-
\frac{\Delta D}{\sqrt{\xi}}~~~.
\label{Q}
\eeq

Of course, being an exact result, \eq{eba} holds to all 
orders in $g$. Comparing Eqs. (\ref{deflam}) and (\ref{xi}), 
we see that
\beq
\xi=\lambda+O(g^2)~~~.
\label{xilam}
\eeq
Thus, to first order in $g$, $Q$ and ${\overline Q}$ of 
\eq{Q} coincide with the eigenvalues of $\w_0^0$ and
$\wb_0^0$ given in \eq{vdo}; conversely, these latter can 
be interpreted as the first-order approximation to the 
exact ones.
 From Eqs.(\ref{xi}) and (\ref{Q}) one reads that the exact 
compactification radius ${\tilde r}$, chemical potential 
${\tilde \mu}$ and Fermi velocity ${\tilde v}$ are given by 
Eqs. (\ref{chempot})-(\ref{comprad}) with $\xi$ in place of
$\lambda$.
Of course, due to \eq{xilam},
${\tilde r}$, ${\tilde \mu}$ and ${\tilde v}$ coincide, respectively,
with $r$, $\mu$ and $v$, to first order in $g$.
It is worthwhile pointing out that all
low-energy effects of the Calogero-Sutherland
interaction are encoded entirely in a unique quantity, namely
the parameter $\xi$,
which in the Bethe Ansatz literature is known as
dressed charge factor \cite{kore}.

Since the exact results can be obtained from the perturbative 
ones simply by changing $\lambda$ into $\xi$, we are led to 
conjecture that the {\it exact} effective hamiltonian of the 
Calogero-Sutherland model $\tilde{\cal H}_{CS}$ is given by 
\eq{hcsf} with $\xi$ in place of $\lambda$.
We may consider this operator as a non-perturbative improvement
of ${\cal H}_{CS}$ which was derived in perturbation theory.

Evidence for the validity of our conjecture, which is certainly
true \cite{kaya} to order $1/N$, is provided
by the calculation of the eigenvalues of ${\tilde {\cal H}}_{CS}$.
In Ref. \cite{flsz} we have checked on several explicit examples 
that these eigenvalues coincide with the exact energy of the 
low-lying excitations given by \eq{eba}.

We conclude by mentioning that the method of 
algebraic bosonization can be applied in principle to any 
(abelian) gapless fermionic hamiltonian consisting of a 
bilinear kinetic term and an arbitrary interaction. 
No special requirements on the form of the dispersion relation 
and the potential are needed. 
In particular, it is not necessary for the system to be integrable.
 
In Ref. \cite{flsz} we have also discussed the algebraic 
bosonization of the Heisenberg model, by mapping it into a 
theory of fermions on a lattice by means of a Jordan-Wigner 
transformation.
Also in this case the spectrum of the low-energy excitations 
can be obtained from the representation theory of the $\winf$
algebra.

\vskip 2.5cm
\noindent
{\large{\bf{Acknowledgments}}}
\vskip 0.5cm
\noindent
We would like to thank the organizers of the Chia meeting, 
and Alberto Devoto in particular, for their kind hospitality.
This research was partially supported by MURST and the EU,
within the framework of the program 
``Gauge Theories, Applied Supersymmetry and Quantum Gravity'',
under contract SCI*-CT92-0789.

\skip 1.5cm


\begin{thebibliography}{99}
\bibitem{cal}   F. Calogero, {\it J. Math. Phys.} {\bf 10} (1969) 2191;
                {\it ibid.} {\bf 10} (1969) 2197; {\it ibid.} {\bf 12}
                (1971) 419.
\bibitem{sut}   B. Sutherland, {\it J. Math. Phys.} {\bf 12} (1971) 246;
                {\it ibid} {\bf 12} (1971) 251;
                \PR{A 4} (1971) 2019; {\it ibid.} {\bf A 5}
                (1972) 1372.
\bibitem{bpz}   A. A. Belavin, A. M. Polyakov and A. B. Zamolodchikov,
                \NP {\bf B 241} (1984) 333; for a review see:
                P. Ginsparg, {\it Applied Conformal Field Theory},
                in {\it Fields, Strings and Critical Phenomena},
                Les Houches School 1988, E. Brezin 
                and J. Zinn-Justin eds.,
                North-Holland, Amsterdam (1990).
\bibitem{hald1} F. D. M. Haldane, \PL{81 A} (1981) 153;
\bibitem{kaya}  N. Kawakami and S.-K. Yang, \PRL{67} (1991)  2493;
                {\it Prog. Theor. Phys. Suppl.} {\bf 107} (1992) 59.
\bibitem{clz}   R. Caracciolo, A. Lerda and G. R. Zemba, \PL{B 352}
                (1995) 304.
\bibitem{flsz}  M. Frau, A. Lerda, S. Sciuto and G. R. Zemba,
                {\it Algebraic bosonization: the study of the Heisenberg
                and Calogero-Sutherland models}, preprint hep-th/9603112.
\bibitem{shen}  I. Bakas, \PL{B 228} (1989) 57;
                C. N. Pope, X. Shen and L. J. Romans, \NP{B 339} (1990);
                for a review see: X. Shen, \IJMP{7 A} (1992) 6953.
\bibitem{kac1}  V. Kac and A. Radul, \CMP{157}(1993) 429.
\bibitem{lutt}  J. M. Luttinger, {\it J. Math. Phys.} {\bf 4}  
                (1963)1154; D. C. Mattis and E. H. Lieb, 
                {\it J. Math. Phys.} {\bf 6} (1965) 304.
\bibitem{solyo} J. S\'olyom, {\it Adv. Phys.} {\bf 28} (1979) 201.
\bibitem{hald} F. D. M. Haldane, {\it ``Luttinger's Theorem and 
                Bosonization 
                of the Fermi surface''}, lectures given at the
                {\it International School of Physics Enrico Fermi},
                Varenna, Italy (July, 1992); {\it J. Phys.} {\bf C 14}
                (1981) 2585. 
\bibitem{kore}  For a recent review see: V. E. Korepin, N. M. Bogoliubov
                and A. G. Izergin, {\it ``Quantum Inverse Scattering
                Method and Correlation Functions''}, Cambridge University
                Press, Cambridge (1993).
\end{thebibliography}
\end{document}